# The role of column density in the formation of stars and black holes

Richard B. Larson

*Department of Astronomy, Yale University, New Haven, CT 06520-8101, USA*

**The stellar mass in disk galaxies scales approximately with the fourth power of the rotation velocity, and the masses of the central black holes in galactic nuclei scale approximately with the fourth power of the bulge velocity dispersion. It is shown here that these relations can be accounted for if, in a forming galaxy with an isothermal mass distribution, gas with a column density above about $8\ M_{\odot}\ \mathrm{pc}^{-2}$ goes into stars while gas with a column density above about $2\ \mathrm{g\ cm}^{-2}$ ($10^4\ M_{\odot}\ \mathrm{pc}^{-2}$) goes into a central black hole. The lower critical value is close to the column density of about $10\ M_{\odot}\ \mathrm{pc}^{-2}$ at which atomic gas becomes molecular, and the upper value agrees approximately with the column density of about $1\ \mathrm{g\ cm}^{-2}$ at which the gas becomes optically thick to its cooling radiation. These results are plausible because molecule formation is evidently necessary for star formation, and because the onset of a high optical depth in a galactic nucleus may suppress continuing star formation and favour the growth of a central black hole.**

The condensation of diffuse matter into stars or black holes requires the loss of both angular momentum and energy, and while angular momentum can be removed or redistributed in a number of ways[1], the thermal energy created by compression must be removed by radiation. The radiative properties of condensing matter are controlled by its optical depth, and this depends on its column density. At one extreme, the most diffuse matter in the universe has a very low column density and cannot form stars because it is kept ionized by the ultraviolet radiation that pervades the universe. At the other extreme, any matter whose column density is so high that radiation is trapped and cooling is prevented cannot easily form stars either, although it can be accreted by an existing object. Between these extremes of column density, the gas can form atoms and molecules that radiate energy efficiently and thus allow normal star formation to occur.

In galaxies, the stellar mass and the black hole mass both increase systematically with the mass of the system, mostly in dark matter. The best-defined correlations are with a characteristic velocity, either the rotation velocity $V$ of a disk or the velocity dispersion $\sigma$ of a bulge, both of which are typically nearly constant. For disk galaxies, the luminosity and inferred stellar mass increase approximately as $V^4$, following the well-known Tully-Fisher relation[2,3,4]. The masses of the central black holes in galaxies with significant bulges increase approximately as $\sigma^4$, being typically about 3 orders of magnitude smaller than the stellar mass[5,6,7]. These relations can be accounted for if there are critical column densities for star formation and black hole formation in galaxies: if the mass distribution in a forming galaxy is approximated by a singular isothermal sphere with a constant velocity dispersion $\sigma$, and if the gas is distributed radially in the same way as the total mass, then as will be shown below, the amount of gas whose column density exceeds any critical value $\Sigma_{\rm crit}$ is proportional to $\sigma^4$ or $V^4$ and inversely proportional to $\Sigma_{\rm crit}$. The scaling relations noted above can then be accounted for if there are two critical column densities in a forming galaxy, such that most of the gas whose column density exceeds a lower critical value goes into stars while most of the gas whose column density exceeds another critical value about 3 orders of magnitude higher goes into a central black hole.

For a singular isothermal sphere with a constant velocity dispersion $\sigma$ and rotation velocity $V$, the mass $M(r)$ inside radius $r$ is $M(r) = V^2r/G = 2\sigma^2r/G$ (ref. 8). If the gas in a forming galaxy is distributed radially like the total mass, and if the ratio of gas mass to total mass is $f_{\rm g}$, the mass of gas inside radius $r$ is $M_{\rm g}(r) = f_{\rm g}M(r)$. If this gas is mostly in a disk, the surface density of gas $\Sigma_{\rm g}(r)$ at any radius $r$ in the disk is $f_{\rm g}\,dM(r)/2\pi r dr$, from which we obtain by differentiation $\Sigma_g(r) = f_g V^2/2\pi Gr = f_g\sigma^2/\pi Gr$. Eliminating $r$ from the expressions for $M_g(r)$ and $\Sigma_g(r)$, the mass of gas inside the radius where the gas surface density is $\Sigma_{crit}$, i.e. the mass of gas with surface density greater than $\Sigma_{crit}$, is

$$M_g(\Sigma_g > \Sigma_{crit}) = f_g^2 V^4/2\pi G^2\Sigma_{crit} = 2 f_g^2 \sigma^4/\pi G^2\Sigma_{crit}. \quad (1)$$

For disk galaxies, the relation between stellar mass and rotation velocity is approximately $M_{stars} \sim 4.5 \times 10^{10} V_{200}^4 M_\odot$ (ref. 4), where $V_{200}$ is the rotation velocity in units of 200 km s$^{-1}$; if the presently observed stars formed from gas whose surface density initially exceeded some critical value $\Sigma_{crit}$, and if $f_g$ is taken to be the cosmic baryon fraction of 1/6, then this relation is reproduced if $\Sigma_{crit} \sim 8\ M_\odot$ pc$^{-2}$. The Tully-Fisher relation can thus be accounted for if, in a simple model of a star-forming disk, all of the gas with surface density above $8\ M_\odot$ pc$^{-2}$ forms stars while all of the gas with lower surface densities remains in gaseous form. For other values of $f_g$, the critical gas surface density scales as $f_g^2$. For the central black holes in galaxies, the relation between black hole mass and bulge velocity dispersion is approximately $M_{BH} \sim 1.5 \times 10^8 \sigma_{200}^4$ (ref. 7), and if we again assume a ratio of gas mass to total mass of $f_g$ = 1/6, as is typically found in starburst galactic nuclei[9], this relation is reproduced with a critical gas surface density $\Sigma_{crit} \sim 2$ g cm$^{-2}$ ($10^4\ M_\odot$ pc$^{-2}$). The relation between black hole mass and bulge velocity dispersion can thus be accounted for if, in the above simple model, all of the gas in a galactic nucleus whose surface density exceeds about 2 g cm$^{-2}$ goes into a central black hole while the gas with lower surface densities goes mostly into stars.

One can ask whether such simple models are realistic enough to be relevant, and whether the derived threshold column densities for star formation and black hole formation are astrophysically plausible. In fact, these models appear to be in reasonable accord with observations, at least on large scales, and the derived threshold column densities correspond in both cases to transitions between different astrophysical regimes that can be expected to be important for star formation and black hole formation.

For disk galaxies, the model assumes that a galaxy acquires most of its mass in a time short compared with the time required for the gas to turn into stars, and that the dark matter quickly settles into an isothermal sphere and the gas into a disk with the same radial mass distribution. Although the innermost parts of large galaxies may be baryon-dominated, the outer parts are not, and it is in these outer parts where the threshold surface density for star formation is determined. For example, in a galaxy like the Milky Way, the threshold gas surface density occurs at a radius of about 25 kpc where dark matter is dominant and the gas and dark matter are not strongly radially segregated. A baryon-to-dark-matter ratio equal to the cosmic value of 1/6 is in reasonable accord with the known properties of the outer disk of the Milky Way galaxy; for example, in the Solar vicinity the disk accounts for about 1/4 of the mass inferred from the rotation curve. A threshold surface density of 8 $M_{\odot}$ pc$^{-2}$ for star formation is plausible astrophysically because it is similar to the column density of about 10 $M_{\odot}$ pc$^{-2}$ above which atomic hydrogen is predicted to become self-shielding against dissociation and hence to become mostly molecular[10]. Observations agree with this prediction, showing that the surface density of atomic hydrogen in galaxies typically saturates at around 10 $M_{\odot}$ pc$^{-2}$ and that gas with higher surface densities is mostly molecular[10]. This threshold surface density for molecule formation also agrees well with the observed threshold gas surface density for star formation in galaxies[11,12], as would be expected if molecule formation is a prerequisite for star formation[13]. The possibility that the Tully-Fisher relation can be accounted for by a threshold gas surface density of ~ 8 $M_{\odot}$ pc$^{-2}$ for star formation therefore appears to receive support from both theory and observations.

The formation and growth of central black holes in galactic nuclei is much less well understood, but an important aspect of the problem has been considered to be the competition between black hole feeding and star formation in a nuclear gas disk[14,15,16]. At radii larger than a few tenths of a parsec, any disk that might feed a black hole at a significant rate is predicted to be gravitationally unstable[14,15], and most of its gas might then go into stars rather than into a central black hole. It is then necessary to understand how star formation might be inhibited and black hole accretion favoured in such a nuclear gas disk. If there were no change in the physics in the central part of a forming galaxy, the same processes that build roughly scale-free distributions of dark matter and stars at larger radii would continue to build them indefinitely toward smaller radii. In models for the formation of individual stars, isothermal gas collapse builds a scale-free density distribution indefinitely toward smaller radii until isothermality is broken by the onset of a high opacity at the centre[17], and we might expect that in the case of galaxy formation, a high opacity will again eventually intervene and change the physics in a dense central region, possibly favouring the formation of a single massive object there.

At the high densities of galactic nuclei, the thermal and radiative properties of the gas are controlled by dust, and cooling is by far-infrared emission from the dust. The dust in the Milky Way nucleus has temperatures of the order of 50 K[18], while the dust in starburst nuclei is typically warmer, with temperatures up to 100 K or more[19]. At a temperature of 50 K the Rosseland mean opacity of star-forming gas is about 0.7 $\mathrm{cm^2\,g^{-1}}$, and at 100 K it is about 3 $\mathrm{cm^2\,g^{-1}}$, remaining in the range of a few $\mathrm{cm^2\,g^{-1}}$ at higher temperatures[20]. The gas in a galactic nucleus therefore becomes optically thick to its cooling radiation at a column density of about 0.3 to 1.5 $\mathrm{g\,cm^{-2}}$, and a column density twice this large is required for the optical depth in the midplane of a gas layer to equal unity. Thus, within a factor of a few, a column density of the order of 1 $\mathrm{g\,cm^{-2}}$ ($5\times10^3\ M_\odot\,\mathrm{pc^{-2}}$) is required for the gas in a galactic nucleus to become optically thick to its cooling radiation.

The column density at which the gas becomes optically thick to its cooling radiation agrees approximately with the column density of $\sim 2\ \mathrm{g\,cm^{-2}}$ above which, in the above model, most of the gas must go into a black hole if the relation between black hole mass and bulge velocity dispersion is to be accounted for. The optical depth of a nuclear gas disk would be expected to play an important role in its evolution because a high optical depth will tend to suppress star formation, and it may also alter the physics in ways that favour black hole accretion. Star formation, or at least the formation of low-mass stars, is likely to be suppressed by a high optical depth because the temperature can then only rise when the gas is compressed; this is true even if the photon diffusion time is short, because an outward temperature gradient is still needed to drive photon diffusion. This is in contrast to normal star formation where efficient cooling to very low temperatures during the early stages plays a crucial role in allowing the gas to fragment to masses of a solar mass or less[21,22]. Even a modest increase in temperature with increasing density can strongly suppress fragmentation[23]. Simulations of star formation in gas captured into orbit around a central black hole in a galaxy show that compressional heating of the optically thick gas near the black hole can indeed yield quite high temperatures and raise the Jeans mass sufficiently that mostly massive stars are formed[24]. The Galactic Center Cluster around the central black hole in our Galaxy does, in fact, have an anomalous stellar mass function that strongly favours massive stars[25].

If the formation of low-mass stars is suppressed, this will reduce the competition from star formation to black-hole feeding. Massive stars may continue to form, but they will not contribute to a population of long-lived stars in a galactic nucleus. A transition from low-mass to high-mass star formation might then be an important first step in the larger transition from star formation to black hole formation in a galactic nucleus. The formation of massive stars in regions of high column density might be expected more generally because a minimum column density of about $1\ \mathrm{g\,cm^{-2}}$ for massive star formation has also been derived from a different argument based on spherical models[26]. While these spherical models differ in many details from the simpler disk models discussed above, in both cases the dust opacity plays an important role and radiative heating inhibits fragmentation into low-mass stars.

Massive stars might also promote black hole accretion by heating the gas in nuclear disks, raising the midplane temperature to $10^3$ K or more in starburst regions[27]. This relatively high temperature may, in addition to suppressing star formation, alter the gas physics in ways that favour black hole accretion. One possibility is that increased coupling to a magnetic field might allow magnetically-driven accretion processes to play a more important role. Another possibility, following directly from the higher temperature, is that pressure disturbances and acoustic waves might play an increased role[28,29]. Even if gravitational torques dominate, as is often assumed, such torques may be enhanced when the disk is heated because a higher temperature means larger-scale, more open spiral patterns and hence stronger torques[28]. Disk heating by massive stars or other heat sources might then not only suppress star formation but promote accretion by a central black hole.

Since substantial heating of a disk, whether by mechanical, magnetic, or radiative means, is possible only if the disk is optically thick, the onset of a high optical depth in a galactic nucleus may lead to a fundamental transition in the astrophysics from dominance by small-scale gravitational processes to more global processes that favour the growth of a central black hole. If so, it may help to explain the relation between black hole mass and bulge velocity dispersion. Also, the fact that the black hole mass in galaxies is typically about 3 orders of magnitude smaller than the stellar mass could readily be understood as following from the fact that the column density of about 1 g cm$^{-2}$ or $5\times10^3$ $M_\odot$ pc$^{-2}$ at which the gas becomes optically thick to its cooling radiation is nearly 3 orders of magnitude higher than the column density of about 10 $M_\odot$ pc$^{-2}$ at which it becomes molecular.

A further prediction is that if star formation is suppressed when the gas surface density exceeds ~ 2 g cm$^{-2}$ or $10^4$ $M_\odot$ pc$^{-2}$, there should be a maximum stellar surface density of this order in galactic nuclei. It has recently been shown[30] that the cores of many different stellar systems have a maximum surface density of order $10^5$ $M_\odot$ pc$^{-2}$, about an order of magnitude higher than the maximum stellar surface density predicted here. Since the critical surface density needed to explain the $M_{\rm BH}$–$\sigma$ relation depends on the square of the gas fraction, it would agree with the observed maximum stellar surface density if a larger gas fraction of 0.5 were assumed, a not implausible value for the early stages of galaxy formation. The same simple model that accounts for the relation between black hole mass and bulge velocity dispersion might then also account for the observed maximum stellar surface density in galaxies.

Since the physics of this simple model is entirely local, it does not depend on any particular scenario of galaxy formation. For example, it can apply even if galaxies are built up from the inside out and if the nucleus and central black hole form first and the outer disks of spirals form later, as long as its assumptions are valid locally at the time of formation . The basic dimensional relations in equation (1) also remain valid even if the mass distribution is not that of an isothermal sphere. The results therefore do not depend sensitively on complex non-local physics such as that involved in feedback effects.

References

1. Larson, R. B. Angular momentum and the formation of stars and black holes. *Rep. Prog. Phys*. (submitted); preprint at ‹http://arxiv.org/abs/0901.4325› (2009).

2. Tully, R. B., & Fisher, J. R. A new method of determining distances to galaxies. *Astron. Astrophys.* **54**, 661–673 (1977).
3. Kassin, S. A., et al. The stellar mass Tully-Fisher relation to $z = 1.2$ from AEGIS. *Astrophys. J.* **660**, L35–L38 (2007).
4. Williams, M. J., Bureau, M., & Cappellari, M. A shared Tully-Fisher relation for spiral and S0 galaxies, in *Galaxy Evolution: Emerging Insights and Future Challenges* (eds Jogee, S., Hao, L., Blanc, G., & Marinova, I.) in press (Astron. Soc. Pacific, San Francisco, 2009); preprint at ‹http://arxiv.org/abs/0902.1500›
5. Kormendy, J., & Richstone, D. Inward bound – the search for supermassive black holes in galactic nuclei. *Annu. Rev. Astron. Astrophys.* **33**, 581–624 (1995).
6. Ferrarese, L., *et al.* A fundamental relation between compact stellar nuclei, supermassive black holes, and their host galaxies. *Astrophys. J.* **644**, L21–L24 (2006).
7. Gültekin, K., *et al.* The $M$–$\sigma$ and $M$–$L$ relations in galactic bulges, and determinations of their intrinsic scatter. *Astrophys. J.* **698**, 198–221 (2009).
8. Binney, J., & Tremaine, S. *Galactic Dynamics: Second Edition* (Princeton University Press, Princeton, 2008).
9. Downes, D., & Solomon, P. M. Rotating nuclear rings and extreme starbursts in ultraluminous galaxies. *Astrophys. J.* **507**, 615–654 (1998).
10. Krumholz, M. R., McKee, C. F., & Tumlinson, J. The atomic-to-molecular transition in galaxies. II: HI and $H_2$ column densities. *Astrophys. J.* **693**, 216–235 (2009).
11. Martin, C. L., & Kennicutt, R. C. Star formation thresholds in galactic disks. *Astrophys. J.* **555**, 301–321 (2001).
12. Schaye, J. Star formation thresholds and galaxy edges: why and where. *Astrophys. J.* **609**, 667–682 (2004).
13. Robertson, B. E., & Kravtsov, A. V. Molecular hydrogen and global star formation relations in galaxies. *Astrophys. J.* **680**, 1083–1111 (2008).
14. Shlosman, I., & Begelman, M. C. Evolution of self-gravitating accretion disks in active galactic nuclei. *Astrophys. J.* **341**, 685–691 (1989).
15. Goodman, J. Self-gravity and quasi-stellar object disks. *Mon. Not. R. Astron. Soc.* **339**, 937–948 (2003).
16. Tan, J. C., & Blackman, E. G. Star-forming accretion flows and the low-luminosity nuclei of giant elliptical galaxies. *Mon. Not. R. Astron. Soc.* **362**, 983–994 (2005).
17. Larson, R. B. The physics of star formation. *Rep. Prog. Phys.* **66**, 1651–1697 (2003).
18. Morris, M., & Serabyn, E. The Galactic Center environment. *Annu. Rev. Astron. Astrophys.* **34**, 645–701 (1996).
19. Solomon, P. M., Downes, D., Radford, S. J. E., & Barrett, J. W. The molecular interstellar medium in ultraluminous infrared galaxies. *Astrophys. J.* **478**, 144–161 (1997).
20. Semenov, D., Henning, Th., Helling, Ch., Ilgner, M., & Sedlmayr, E. Rosseland and Planck mean opacities for protoplanetary disks. *Astron. Astrophys.* **410**, 611–621 (2003).
21. Larson, R. B. Thermal physics, cloud geometry and the stellar initial mass function. *Mon. Not. R. Astron. Soc.* **359**, 211–222 (2005).

22. Jappsen, A.-K., Klessen, R. S., Larson, R. B., Li, Y., & Mac Low, M.-M. The stellar mass spectrum from non-isothermal gravoturbulent fragmentation. *Astron. Astrophys.* **435**, 611–623 (2005).
23. Li, Y., Klessen, R. S., & Mac Low, M.-M. The formation of stellar clusters in turbulent molecular clouds: effects of the equation of state. *Astrophys. J.* **592**, 975–985 (2003).
24. Bonnell, I. A., & Rice, W. K. M. Star formation around supermassive black holes. *Science* **231**, 1060–1062 (2008).
25. Bartko, H., *et al.* An extremely top-heavy IMF in the Galactic Center stellar disks. *Astrophys. J.* (submitted); preprint at‹http://arxiv.org/abs/0908.2177› (2009).
26. Krumholz, M. R., & McKee, C. F. A minimum column density of 1 g $cm^{-2}$ for massive star formation. *Nature* **451**, 1082–1084 (2008).
27. Thompson, T. A., Quataert, E., & Murray, N. Radiation pressure-supported starburst disks and active galactic nucleus fueling. *Astrophys. J.* **630**, 167–185 (2005).
28. Larson, R. B. The evolution of protostellar disks, in *The Formation and Evolution of Planetary Systems* (eds Weaver, H. A., & Danly, L.) 31–54 (Cambridge University Press, Cambridge, 1989).
29. Larson, R. B. Non-linear acoustic waves in discs. *Mon. Not. R. Astron. Soc.* **243**, 588–592 (1990).
30. Hopkins, P. F., Murray, N., Quataert, E., & Thompson, T. A. A maximum stellar surface density in dense stellar systems. *Mon. Not. R. Astr. Soc.* (submitted); preprint at ‹http://arxiv.org/abs/0908.4088›